\documentclass[12pt]{article}

\usepackage{epsfig}
\begin{document}
\begin{flushright}

DTP--00/57\\ hep-th/0007214 \\

\end{flushright}

\newcommand{\Title}[1]{{\baselineskip=26pt \begin{center}
            \Large   \bf #1 \\ \ \\ \end{center}}}
\newcommand{\Author}{\begin{center}\large \bf
            M.\, Ablikim$^a$ and E.\, Corrigan$^b$\end{center}}
\newcommand{\Address}{\begin{center} \it
            $^a$  Department of Mathematical Sciences, University of
            Durham,\\ Durham, DH1 3LE, UK\\
            $^b$ Department of Mathematics, University of York,
            \\Heslington, York YO10 5DD, UK
      \end{center}}

\bigskip
\bigskip
\bigskip

\baselineskip=20pt

\Title{On the perturbative expansion of boundary reflection
factors of the supersymmetric sinh-Gordon model}
\Author

\Address
\vspace{2cm}

\begin{abstract}

The supersymmetric  sinh-Gordon model on a half-line with integrable boundary
conditions is considered perturbatively  to verify conjectured exact
reflection factors to one loop order. Propagators for the boson and fermion
fields restricted to a half-line contain several novel features and are
developed as prerequisites for the calculations.

\end{abstract}

\section{Introduction}
\setcounter{equation}{0}

Given an interesting field theory, it is traditional to develop and examine
its supersymmetric extensions. In four dimensions, supersymmetric field
theories provide the prime examples of situations in which quantities of
physical interest may be calculated exactly. For this reason they are an
important source of ideas and intuition. In theories of strings and branes,
supersymmetry is more or less mandatory. However, in two dimensions, for
nonlinear models, the two requirements of supersymmetry and integrability do
not always sit easily together. There are many models which are entirely
bosonic and yet interesting quantities may be calculated exactly, often by
indirect algebraic means which are not available for use in four dimensions;
there are others which contain fermions and are integrable, and yet not
supersymmetric. On the other hand, there are many examples of two-dimensional
models which are both integrable and supersymmetric; for a  selection see
~\cite{Ol,Evans1,Evans2,Papa,PZ1}. The  supersymmetric version of the
sine-Gordon model was introduced many years ago by Hruby~\cite{Hruby}, and di
Vecchia and Ferrara~\cite{Ferrara1}; Shankar and Witten \cite{SW} constructed
its exact S-matrix  which was subsequently further explored by Schoutens
\cite{Schoutens} and Ahn \cite{Ahn1}.

If a field theory is restricted to a half-line by integrable
boundary conditions then it turns out that supersymmetry is
further constrained, and more restrictive. For example,
Warner~\cite{W1,W2} discussed quantum
integrable models possessing $N=2$ supersymmetry and concluded that
 only half the supersymmetry may be retained in the presence of the
 boundary.  In the case of the sine-Gordon model, it
was pointed out by Inami, Odake and Zhang~\cite{IOZ} that only two isolated
boundary conditions are compatible with both supersymmetry and integrability.
This is a striking and surprising result since without supersymmetry Ghoshal
and Zamolodchikov~\cite{GZ}  had earlier pointed out that there should be a
two parameter family of nonlinear boundary conditions compatible with
integrability. More recently,  using general arguments,  exact reflection
matrices for the breathers and their fermionic partners within the $N=1$
supersymmetric sine-Gordon theory have been conjectured~\cite{AhnKoo, MS1}.

In this paper, we will examine supersymmetric  sinh-Gordon
theory  restricted to a half-line by  integrable boundary
conditions. We will argue that the results found by Moriconi
and Schoutens need to be adjusted slightly to agree with the
classical limit and with the lowest orders of perturbation theory
in the sinh-Gordon coupling constant.
In order to carry out a low order check in perturbation theory
we will also need to construct propagators for both the boson and
the fermion. The boson propagator was constructed before~\cite{Corrigan1}
but the fermion propagator is constructed here for the first time.

The paper is organised as follows:
in section two, we  summarise the main features of the model;
the boson and fermion propagators are described in section three
together with a brief discussion on the r\^ole played by bound
states in the linearised theory;
the construction of the supersymmetric scattering matrices and the
conjectured reflection factors for the two allowed
boundary conditions are presented in section four together with
reasons for deviating from the suggestions made by Moriconi
and Schoutens; in the subsequent sections we develop the
perturbation theory and check that the fermion reflection factors agree
with the perturbation expansion up to second order in the
bulk coupling constant. The final section is reserved for
concluding remarks.

\section{The supersymmetric sinh-Gordon model with one boundary}
\setcounter{equation}{0}

To establish the conventions we shall use, it is convenient to  start with the
supersymmetric sinh-Gordon model in the bulk  described by the Lagrangian
density
\begin{equation}
  {\cal L} = \frac12 \partial_\mu \phi \partial^\mu \phi -
   \frac{m^2}{2\beta^2}
 \cosh\sqrt2\beta \phi - \bar\psi i\gamma^\mu \partial_\mu \psi
  + m\bar\psi \psi\cosh\frac{\beta\phi}{\sqrt2},\label{Lag}
\end{equation}
where $\beta$ is a real coupling constant, $m$ is a mass parameter, $\phi$ is
a real scalar field and $\psi$ is a two-component Majorana fermion.   We
choose the $\gamma$-matrices to be purely imaginary   represented by
\begin{eqnarray}
  \gamma^0 = \left( \begin{array}{c c} 0 & -i \\ i & 0 \\ \end{array}
  \right)   \qquad
  \gamma^1 = \left( \begin{array}{c c} 0 & i \\ i & 0 \\
  \end{array}
  \right). \label{rep}
\end{eqnarray}
With this choice,  the charge-conjugation matrix is unity and a Majorana
spinor has two real components.

The Lagrangian~(\ref{Lag}) is invariant under the following set of
supersymmetry transformations:
\begin{eqnarray}
&&\delta \phi = \bar\epsilon \psi + \bar\psi \epsilon, \nonumber \\
&&\delta \psi = \left( i\gamma^\mu \partial_\mu \phi +
   \frac{\sqrt2 m}{\beta}\sinh\frac{\beta\phi}{\sqrt2} \right)
   \epsilon, \label{susy} \\
&&\delta \bar\psi = \bar\epsilon \left( -i\gamma^\mu \partial_\mu \phi +
   \frac{\sqrt2 m}{\beta}\sinh\frac{\beta\phi}{\sqrt2}\right),  \nonumber
\end{eqnarray}
where $\epsilon$ is a constant  Majorana spinor which anticommutes  with
$\psi$. From the Lagrangian~(\ref{Lag}) the field equations are:
\begin{eqnarray}
 &&\partial^2 \phi = -\frac{m^2}{\sqrt2 \beta} \sinh \sqrt2 \beta
\phi - \frac{m}{\sqrt2}\bar\psi \psi \sinh\frac{\beta\phi}{\sqrt2},
 \nonumber\\ \label{eqm}
  &&i\gamma^\mu \partial_\mu \psi =  m\psi \cosh\frac{\beta\phi}
{\sqrt2}, \\
  &&i\partial_\mu\bar\psi\gamma^\mu = - m\bar\psi
    \cosh\frac{\beta\phi}{\sqrt2}. \nonumber
\end{eqnarray}

The model may be restricted to a half-line   by adding a boundary term to the
action which enforces an
 integrable boundary
condition on the fields. For convenience, we shall take the boundary to be
situated at $x=0$. Following the arguments of Inami, Odake and Zhang, the
action for the sinh-Gordon model on a half-line may be written as follows:
\begin{eqnarray}\label{susyLag}
 S &=& \int_{-\infty}^\infty dt  \int_{-\infty}^0 dx \left[
\frac{1}{2} \partial_\mu \phi \partial^\mu \phi - \frac{m^2}{2\beta^2}
\cosh\sqrt2\beta \phi - \bar\psi i\gamma^\mu \partial_\mu \psi + m
\bar\psi
\psi
\cosh\frac{\beta \phi}{\sqrt2} \right] \nonumber \\
   & &\hspace*{3cm} -\int_{-\infty}^\infty dt \left [\pm
\frac{2m}{\beta^2} \cosh \frac{\beta \phi}{\sqrt2}
\mp\frac{1}{2}\bar\psi \psi
\right].
\end{eqnarray}
The action including the boundary part is supersymmetric if and only if
the two components of the
parameter $\epsilon$ satisfy
\begin{equation}
      \epsilon_1 = \mp \epsilon_2,
\end{equation}
which confirms that only half  the supersymmetry of the bulk theory is
preserved in the presence of  boundary conditions.

In addition to the bulk equations of motion~(\ref{eqm}) in the region $x<0$,
the boundary conditions for the fields at $x=0$ follow from (\ref{susyLag})
and are
\begin{equation}\label{boundary1}
   \partial_x \phi =  \mp\frac{\sqrt2m}{\beta} \sinh \frac{\beta
    \phi}{\sqrt 2} , \quad
\psi_1 = \pm \psi_2 .
\end{equation}
We shall refer to these two boundary conditions as $\cal{BC}^{\pm}$, the $\pm$
corresponding to the signs relating the fermion components in each of the two
cases given in (\ref{boundary1}).

\section{Boson and fermion propagators}
\setcounter{equation}{0}

The construction of the boson propagator for the sinh-Gordon model in the
presence of integrable boundary conditions was given in~\cite{Corrigan1}. In
the supersymmetric case we have just two kinds of boundary condition
preserving both supersymmetry and integrability. The boson propagators
corresponding to these are given by
\begin{equation}\label{bosonpropagator}
G^{\pm}(x,t;x',t')=\int \frac{d\omega}{2\pi} \int \frac{dk}{2\pi} \;
\frac{ie^{-i\omega(t-t')}}{\omega{}^2 -
  k{}^2 - m^2 +i\epsilon}\left[ e^{ik(x -x')} + K_b^{\pm}(k) \;
e^{-ik(x + x')} \right].
\end{equation}
The coefficients of the reflected term in the integrand of
(\ref{bosonpropagator}) correspond to the `classical'
reflection factors of the model linearised about the ground
state solution $\phi=0$,
\begin{equation}
K_b^{\pm} (k) = \frac{ik \pm m}{ik \mp m}=\frac{i\sinh\theta\pm 1}
{i\sinh\theta\mp 1}. \label{cr1}
\end{equation}
In (\ref{cr1}), the second form of the expression refers to the on-shell
reflection factor for a particle with rapidity $\theta$ for which
$k=m\cosh\theta$.

One point which was not emphasised in \cite{Corrigan1} but which is important
here is the fact that for the boundary condition $\cal{BC}^-$ there is a
boundary bound state even in the linearised theory. This reveals itself when
calculating $(\partial^2+m^2)G^{\pm}$. The first term in the integrand of
(\ref{bosonpropagator}) leads to the usual bulk contribution but the second,
which can be evaluated by closing the $k$-contour in the upper half-plane
(since $x+x^\prime<0$), will lead to an extra piece arising from the
additional pole in $K_b^-$ at $k=im$. Since there is a normalizeable field
configuration corresponding to the bound state field configuration $\phi\sim
e^{mx}$ (which has zero frequency and decays exponentially for $x<0$), there
should have been an additional contribution to the propagator for the case
$\cal{BC}^-$ which would  effectively remove the unwanted pole.

To examine this a little more, consider first a linear scalar field theory
with mass parameter $m$ and  boundary condition
\begin{equation}
\partial_x\phi =-\lambda \phi,
\end{equation}
at $x=0$.
There is  a bound state solution $\phi\sim e^{-i\omega t} e^{-\lambda x}$
provided $-m<\lambda <0$ and $\omega^2=m^2-\lambda^2$. In this situation, the
free boson propagator takes the form (\ref{bosonpropagator}) with
\begin{equation}
K_b (k) = \frac{ik +\lambda}{ik -\lambda},
\end{equation}
together with an extra piece
\begin{equation}\label{extrabosonpiece}
-2\lambda i \int\frac{d\omega}{2\pi}\,\,  \frac{e^{-i\omega (t-t^\prime )}
e^{-\lambda(x+x^\prime)}}{\omega^2+\lambda^2-m^2+i\epsilon},
\end{equation}
constructed from the normalised bound state field. The `$i\epsilon$'
prescription has the usual meaning: that is,  the contour of
$\omega$-integration passes below the pole at $\omega=\sqrt{m^2-\lambda^2}$
and above the pole at $\omega=-\sqrt{m^2-\lambda^2}$. It is easy to check that
the additional piece serves to remove the effect of the extra pole in $K_b$ at
$k=-i\lambda$. For the supersymmetric sinh-Gordon model, $\lambda =-m$, and
the additional piece is not defined in the $\epsilon\rightarrow 0$ limit.
However, neither is the right-hand side of (\ref{bosonpropagator}) because of
the pole at $k=im$. On the other hand, happily, the two components
(\ref{bosonpropagator}) and (\ref{extrabosonpiece}), taken together,  are
well-defined. Thus, in this article, we shall adopt a pragmatic approach which
uses the expression (\ref{bosonpropagator}) but ignores the pole at $k=im$
wherever it occurs.

Next, let us consider the fermion propagator. We are familiar with the usual
expression for a  fermion propagator on the whole line. In two dimensions,
with our choice of $\gamma$-matrices, it would be written
\begin{equation}
  S_F(x-x')= \int \frac{d\omega}{2\pi} \frac{dk}{2\pi}\;
    \frac{i e^{-i\omega (t -t')}}{\omega{}^2 - k{}^2 - m^2 +
    i\epsilon} \left(
    \begin{array}{cc} m & -i(\omega +k) \\ i(\omega -k) & m \\
    \end{array} \right) e^{ik(x -x')}.
\end{equation}

In the presence of the boundary we need to modify the standard fermion
propagator, ensuring  not only that it performs as a propagator in the bulk
but also  that it respects the fermion part of the boundary
conditions~({\ref{boundary1}). Clearly, the usual relationship between  boson
and fermion propagators will no longer hold. Bearing this in mind, our
expression for the fermion propagators is the following:
\begin{eqnarray}\label{fermionpropagator}
  S^{\pm}_F(x,t; x',t') &=& \int \frac{d\omega}{2\pi} \frac{dk}{2\pi}\;
    \frac{i e^{-i\omega (t -t')}}{\omega{}^2 - k{}^2 - m^2 +
    i\epsilon} \left[ \left(
    \begin{array}{cc} m & -i(\omega +k) \\ i(\omega - k) & m \\
    \end{array} \right) e^{ik(x -x')} \right. \label{fermion1} \nonumber\\
  &&\hspace*{3cm} \left. \pm \frac{\omega}{ik \mp m} \left(
   \begin{array}{cc} \omega - k & -im \\ im & \omega + k \\
   \end{array} \right) e^{-ik(x + x')} \right].
\end{eqnarray}
As far as the boundary conditions are concerned it is not difficult
to verify that at $x=0$,
\begin{equation}
\left(S_F^{\pm}\right)_{1\ a}=\pm \left(S_F^{\pm}\right)_{2\ a}, \quad
a=1,2.
\end{equation}
However, a calculation of $(-i\gamma\cdot\partial +m)S_F^{\pm}$, while giving
the expected result for $S_F^{+}$ (because the second term in the integrand
integrates to zero on closing the contour in the upper half $k$-plane),
reveals an additional contribution for $S_F^{-}$ owing to the extra pole at
$k=im$. This pole reflects the fact that there is also a fermion bound state
for $\cal{BC}^{-}$ (as there should be because of the supersymmetry) and that
therefore the expression for the propagator $S_F^{-}$ requires adjustment.
Again, our pragmatic approach amounts to ignoring the extra pole and using
(\ref{fermionpropagator}) without alteration.

\phantom{}From the expression (\ref{fermionpropagator}), it is natural to take
the `classical' fermion reflection factors to be given by
\begin{equation}
 K_f^{\pm}= \pm \frac{\omega}{ik \mp m}=
 \pm\frac{\cosh\theta}{i\sinh\theta\mp 1},  \label{cr2}
\end{equation}
and, as before, the second expression refers to the the on-shell reflection
factors which are related to the free bosonic reflection factors by,
\begin{equation}
K_f^{\pm}=\pm i\sqrt{K_b^{\pm}}.
\end{equation}

\section{The construction of the $S$-matrix and the reflection factors for
  the supersymmetric theory}
\setcounter{equation}{0}

An $N=1$ supersymmetric theory contains a conserved Majorana supercharge. In
terms of the chiral components $Q_{\pm}$ of the supercharge we can write the
on-shell supersymmetry algebra as follows,
\begin{equation}
   Q_{\pm}^2 = me^{\pm \theta}, \quad \quad \{Q_+, Q_-\} = 0, \quad \quad
           \{Q_L, Q_{\pm}\} = 0, \label{algebra}
\end{equation}
where the operator $Q_L$  has eigenvalue $+1$ on bosonic states and $-1$ on
fermionic states. The single particle states of a massive supersymmetric
theory contain either one boson or one fermion of  mass $m$, and will be
denoted by $|b(\theta)>$ or $|f(\theta)>$ respectively.

It follows from the algebra~(\ref{algebra}) that the action of the
supercharges $Q_{\pm}$ on the one  particle states can
be represented by
\begin{eqnarray}
  & Q_+|b(\theta)> = \sqrt m e^{\theta/2} |f(\theta)>, \quad
   & Q_+|f(\theta)> = \sqrt m e^{\theta/2} |b(\theta)>,\nonumber\\
  & Q_-|b(\theta)> = i \sqrt m e^{-\theta/2} |f(\theta)>, \quad
   & Q_-|f(\theta)> = -i\sqrt m e^{-\theta/2} |b(\theta)>,
\end{eqnarray}
corresponding to a  realization of~(\ref{algebra}) in terms of
the Pauli matrices
\begin{equation}
    Q_+ = \sqrt m e^{\theta/2}\left(
\begin{array}{c c} 0 & 1 \\
                   1 & 0 \\
\end{array}                  \right), \quad
    Q_- = \sqrt m e^{-\theta/2}\left(
\begin{array}{c c} 0 & -i \\ i & 0 \\
          \end{array} \right), \quad
Q_L = \left( \begin{array}{c c} 1 & 0 \\ 0 & -1 \\
          \end{array} \right).    \label{4.3}
\end{equation}

The $S$-matrix in the supersymmetric
theory is tightly constrained by
the supersymmetry and has been given by Schoutens~\cite{Schoutens}
in the following form:
\begin{equation}
  S=S_b(\Theta) \; S_{s}(\Theta).
\end{equation}
Here, $S_b$ is the  $S$-matrix for the sinh-Gordon model without fermions
\cite{Fadd78},
 $S_{s}$ is  the supersymmetric
part, responsible for mixing bosons and fermions, and $\Theta$ is the rapidity
difference of the scattering particles. In detail, using a
 shorthand `bracket' notation \cite{Bra90} we have
\begin{equation}\label{shGSmatrix}
S_b(\Theta)= -\frac{1}{(B)(2-B)},\qquad (x)=\frac{\sinh\left(\frac{\Theta}{2}
+\frac{i\pi x}{4}\right)}{\sinh\left(\frac{\Theta}{2}
-\frac{i\pi x}{4}\right)}
\end{equation}
where the coupling constant enters via
\begin{equation}
B(\beta)=\frac{1}{2\pi} \; \frac{\beta^2}{1+\beta^2/4\pi},
\end{equation}
and,
\begin{eqnarray}
  S_{s}(\Theta) &=& f(\Theta) \left(\begin{array}{c c c c}
                1-\tanh^2\frac{\Theta}{4} & 0 & 0 &
-2i\tanh\frac{\Theta}{4}\\
               0 & 0 & 1+\tanh^2\frac{\Theta}{4} & 0 \\
               0 & 1+\tanh^2\frac{\Theta}{4} & 0 & 0\\
               -2i\tanh\frac{\Theta}{4} & 0 & 0 &
1-\tanh^2\frac{\Theta}{4} \\
                  \end{array} \right)\nonumber \\
                   && \nonumber \\
              && \ \ \ \ \ \ \ \ \ \ \ \ \ \ \ \ \ \ \ \ \ \
               + g(\Theta) \left(\begin{array}{c c c c}
                1 & 0 & 0 & 0\\
                0 & 1 & 0 & 0 \\
                0 & 0 & 1 & 0\\
                0 & 0 & 0 & -1 \\
                  \end{array} \right), \label{S}
\end{eqnarray}
where
\begin{eqnarray}
 && f(\Theta) = f_0\frac{\left(\cosh\frac{\Theta}{2}
+ 1 \right)}{\sinh\Theta} \; g(\Theta),\\
&& \nonumber \\
 && g(\Theta) = \frac{\sinh\frac{\Theta}{2}}
                        {\sinh\frac{\Theta}{2} - i\sin \rho\pi}
\;
                     \exp \left[-i \int^\infty_0 \frac{dt}{t}
                   \; \frac{\sinh \rho t \;
   \sinh \left( 1+ \rho \right)t}
        {\cosh^2 \frac{t}{2}\cosh t} \; \sin\frac{\Theta t}{\pi} \right].
\end{eqnarray}
The parameter $\rho$ depends on the coupling $\beta$ and we expect
$\rho =B/4$ in our conventions.

 Moriconi and Schoutens \cite{MS1} assumed
that the reflection matrix can be factorised  similarly and will therefore
take the form,
\begin{equation}\label{Rdef}
  R(\theta)=R_b(\theta) \; R_{s}(\theta). \label{rf}
\end{equation}
Here, $\theta$ is the rapidity of the reflecting particle,
$R_b(\theta)$ would be the reflection matrix for the bosonic part of the
theory in the absence of fermions, and $R_{s}$ is the supersymmetric
part.
Assuming further that the boundary does not convert bosons to fermions,
or vice-versa, supersymmetry constrains $R_{s}$ to have the form
\begin{eqnarray}\label{Rsdef}
  R_{s}^{\pm}(\theta) =  Z^{\pm}(\theta) \left(\begin{array}{c c}
    \cosh(\frac{\theta}{2} \pm \frac{i\pi}{4}) & 0\\
    0 & \cosh(\frac{\theta}{2} \mp \frac{i\pi}{4})\\
  \end{array} \right).
\end{eqnarray}
Thus, we may write equivalently:
\begin{eqnarray}
K_b^{\pm}(\theta )&=&R_b^{\pm}(\theta )Z^{\pm}(\theta)
\cosh\left(\frac{\theta}{2} \pm \frac{i\pi}{4}\right)\nonumber\\
K_f^{\pm}(\theta )&=&R_b^{\pm}(\theta )Z^{\pm}(\theta)
\cosh\left(\frac{\theta}{2} \mp \frac{i\pi}{4}\right).
\end{eqnarray}
It is interesting to notice that the ratios of boson to
fermion reflection factors do not depend on anything other
than the rapidity. In fact from (\ref{Rdef}) and (\ref{Rsdef}) we
deduce,
\begin{equation}
     \frac{K_b^{\pm}(\theta)}{K_f^{\pm}(\theta)} =
\frac{\cosh(\frac{\theta}{2} \pm \frac{i\pi}{4})}{\cosh(\frac{\theta}{2}
\mp \frac{i\pi}{4})}=
\frac{1 \pm i\sinh \theta}
            {\cosh\theta},\label{4.18}
\end{equation}
which is in perfect agreement with the ratios of the classical reflection
factors given in (\ref{cr1}) and (\ref{cr2}). In the classical limit
the complete reflection matrix must match the boson and fermion
classical factors. This requires particular classical limits for
$Z^{\pm}(\theta)$, namely
\begin{equation}\label{Zlimits}
Z^{\pm}(\theta)\rightarrow \frac{1}{\cosh\left(\frac{\theta}{2}\pm
\frac{i\pi}{4}\right)}.
\end{equation}
In addition, the factor $Z^{\pm}(\theta)$ is constrained by the requirements
of unitarity
\begin{equation}
R(\theta)R(-\theta)=1,
\end{equation}
and by boundary crossing unitarity \cite{GZ,FK}
\begin{equation}\label{boundarycrossing}
1=\sum_{c=b,f}K_b\left(\theta-\frac{i\pi}{2}\right)S_{cc}^{bb}(2\theta)
K_c\left(\theta+\frac{i\pi}{2}\right),
\end{equation}
which must be satisfied by the full reflection factor.
Because of the factorised
forms of both the S-matrix and the reflection factor, and the fact that
we shall choose the bosonic part of the reflection factor to satisfy
(\ref{boundarycrossing}) in conjunction with the sinh-Gordon S-matrix
(\ref{shGSmatrix}),
these conditions lead to
\begin{eqnarray}\label{Zmess}
 && Z^{\pm}(\theta)Z^{\pm}(-\theta) = 2/\cosh\theta,\label{3.15}\\
 &&\nonumber \\
 && \frac{Z^{\pm}\left( \frac{i\pi}{2}-\theta \right)}{Z^{\pm}\left(
  \frac{i\pi}{2}+ \theta \right)} = \mp S_{bb}^{bb}(2\theta) +
 i\left(\coth\frac{\theta}{2}\right)^{\pm 1} \; S_{ff}^{bb}
 (2\theta),
\end{eqnarray}
where   $S_{bb}^{bb}(2\theta)$ and $S_{ff}^{bb} (2\theta)$ should be extracted
from~(\ref{S}).

Given the classical limits (\ref{Zlimits}) it is natural to set
\begin{equation}
Z^{\pm}=\frac{\tilde{Z}^{\pm}}{\cosh\left(\frac{\theta}{2}\pm
\frac{i\pi}{4}\right)},
\end{equation}
in which case,
\begin{eqnarray}\label{Zz}
&& \tilde{Z}^{\pm}(\theta)\tilde{Z}^{\pm}(-\theta) = 1\nonumber\\ &&\\
&&\frac{\tilde{Z}^{\pm}\left( \frac{i\pi}{2}-\theta \right)}{
\tilde{Z}^{\pm}\left(
  \frac{i\pi}{2}+ \theta \right)} = \frac{\sinh\theta \mp 2f_0}
  {\sinh\theta -i\sin \rho\pi}\, \exp \left[-i \int^\infty_0 \frac{dt}{t}
                   \; \frac{\sinh \rho t \;
   \sinh \left( 1+ \rho \right)t}
        {\cosh^2 \frac{t}{2}\cosh t} \; \sin\frac{\theta t}{\pi} \right].\nonumber
\end{eqnarray}
Then the solutions we want will satisfy
\begin{equation}
\tilde{Z}^{\pm}(\theta)\rightarrow 1
\end{equation}
in the classical limit. Clearly,  $f_0=\pm (i/2)\sin \rho\pi$, and we shall
take $f_0=-(i/2)\sin \rho\pi$.

The equations (\ref{Zz}) are solved by:
\begin{equation}\label{zminus}
      \tilde{Z}^-\left( \theta \right)=  \exp \left[ \frac{i}{2} \int^\infty_0
    \frac{dt}{t} \; \frac{\sinh \rho t \ \sinh \left( 1
+\rho \right)t}{\cosh^2 \frac{t}{2} \cosh^2 t}
\sin \frac{2 \theta t}{\pi} \right],
\end{equation}
and
\begin{eqnarray}\label{zplus}
 \tilde{Z}^+\left( \theta \right) &=&
    \exp \left[- 2i \int^\infty_0 \frac{dt}{t} \;
     \frac{\sinh \frac{\rho t}{2} \sinh\frac12\left(
     1+\rho\right) t}{\cosh^2 \frac{t}{2}}
       \sin \frac{\theta t}{\pi} \right] \nonumber \\
   &&\hspace*{2.6cm} \exp \left[ \frac{i}{2}
   \int^\infty_0 \frac{dt}{t} \; \frac{\sinh \rho t \
\sinh
  \left( 1+\rho \right)t}{\cosh^2 \frac{t}{2} \cosh^2 t}
  \sin \frac{2 \theta t}{\pi} \right].
\end{eqnarray}
Notice that these are not quite the same as the proposals made in \cite{MS1}
since Moriconi and Schoutens took the view that the classical limit of
a free boson reflection factor should be unity; an assumption which is not
generally valid, as we have seen.

Ghoshal~\cite{Ghoshal} has calculated a formula for the quantum reflection
matrix for the breather states of the sine-Gordon model. The reflection factor
for the sinh-Gordon model is presumed to be deduced from the lightest breather
reflection factor in the sine-Gordon theory by analytic continuation in the
coupling constant (replacing $\beta $ by $i\beta$), leading to the expression
\begin{equation}
   R_b\left( \theta \right)=\frac{\left( 2-B/2 \right)\left( 1\right)
 \left( 1+B/2 \right)}
   {\left( 1-E \right) \left(
    1+E \right)\left( 1-F \right)\left( 1+F \right)}, \label{shg}
\end{equation}
where the coupling dependent functions $E$ and $F$ also depend on the boundary
parameters introduced via the boundary potential.\footnote{$E$ and $F$ are
related to the  parameters $\eta$ and $\vartheta$ in Ghoshal's notation by
$$E=\frac{\eta}{\pi}B, \quad \quad F=\frac{i\vartheta}{\pi}B.$$ }
In the supersymmetric theory, we consider the boundary
conditions~(\ref{boundary1}) for which $F=0$. On the other hand, an expression
for $E$ has been found recently by comparing two independent calculations of
the boundary breather spectrum \cite{Corrigan2}. This translates in the
present situation with two  possible boundary conditions (\ref{boundary1}) to
\begin{equation}
 {\cal BC}^{+}: \quad E=0, \quad {\cal BC}^-:\quad  E = 2(1-B/2).
 \label{conjecture}
\end{equation}
Thus we take
\begin{equation}\label{Rbplus}
R_b^+= \frac{\left(1+B/2\right)\left(2-B/2\right)}{(1)^3},
\end{equation}
and
\begin{equation}\label{Rbminus}
R_b^-=\frac{\left(1+B/2\right)\left(2-B/2\right)(1+B)(1-B)}{(1)}.
\end{equation}
Notice that (\ref{Rbminus}) contains the bound-state pole (in the factor
$(1-B)$) at $\theta =i(1-B)\pi/2$, whereas (\ref{Rbplus}) contains no bound
states. The suggestions made by Moriconi and Schoutens were different but for
comparison we list them here:
\begin{equation}\label{MSRbplus}
R_b^+= \frac{\left(2-B/2\right)}{\left(1+B/2\right)(1)},
\end{equation}
and
\begin{equation}\label{MSRbminus}
R_b^-=(1)\left(1+B/2\right)\left(2-B/2\right),
\end{equation}
corresponding to $E=B/2$ and $E=2$, respectively. One could argue that the
fermions ought to modify the conclusions reached in \cite{Corrigan2} in such a
manner as to prevent any renormalisation of the position of the bound state
pole. In that sense, (\ref{MSRbminus}) would seem to be a better guess since
the boundary bound state retains its position at $\theta =i\pi/2$. We note
too, that the two expressions (\ref{Rbplus}) and (\ref{MSRbminus}) share the
property of being invariant under the duality transformation $B\rightarrow
2-B$. However, we are not sure too much should be read into this since the
other factors $Z^\pm (\theta )$ do not share the same property. An appeal to
lowest order perturbation theory does not help either since to order $\beta^2$
we have identical expansions for (\ref{Rbplus}) and (\ref{MSRbplus}),
\begin{equation}\label{Rbplusexpansion}
R_b^+\sim \left(\frac{i\sinh \theta+1}
    {i\sinh \theta-1}\right)\left[ 1- \frac{i \beta^2}{8} \sinh \theta
  \left( \frac{1}{\cosh \theta+1}-\frac{1}{\cosh \theta} \right)
   \right],
\end{equation}
as indeed we do for (\ref{Rbminus}) and (\ref{MSRbminus}),
\begin{equation}\label{Rbminusexpansion}
R_b^-\sim \left(\frac{i\sinh \theta-1}
    {i\sinh \theta+1}\right)\left[ 1- \frac{i \beta^2}{8} \sinh \theta
  \left( \frac{1}{\cosh \theta+1}-\frac{1}{\cosh \theta} \right)
   \right].
\end{equation}
It should be possible to distinguish between (\ref{Rbminus})
and (\ref{MSRbminus}) by following the semi-classical quantization
of the boundary breathers taking the fermions into account. However,
this analysis has not yet been carried out.

To conclude this section we shall prepare the way for comparing the
reflection factors with low order perturbation theory by giving their expansions
to order $\beta^2$. This is straightforward apart from a couple of
complicated integrals arising from the $Z$-factors.
For example, setting $\rho\sim \rho_0 \beta^2/8\pi$, we have
\begin{eqnarray}\label{Zplusexpansion}
    \tilde{Z}^+\left( \theta \right) &\sim&
  1 - \rho_0 \,\frac{i\beta^2}{16\pi} \left[
2\int^\infty_0 dt\;
\frac{\sinh\frac{t}{2}}{\cosh^2\frac{t}{2}}\sin\frac{t\theta}
    {\pi}%\nonumber\\ &&\hspace*{3cm}
- \int^\infty_0 dt \;
\frac{\sinh t}{\cosh^2\frac{t}{2}\cosh^2t}
\sin\frac{2t\theta}{\pi}\right]\nonumber\\
&&\nonumber\\
&=&1-\rho_0 \frac{i\beta^2}{16\pi} \;
 \left[ \frac{2\theta}{\cosh\theta} -
 \pi\sinh\theta\left(\frac{1}{\cosh\theta + 1}
  -\frac{1}{\cosh\theta} \right)\right],
\end{eqnarray}
where we have used the following two facts~\cite{GR}:
\begin{eqnarray}
\int^\infty_0 dt\;
\frac{\sinh\frac{t}{2}}{\cosh^2\frac{t}{2}}\sin\frac{t\theta}
    {\pi} &=& \frac{2\theta}{\cosh\theta}\nonumber\\
&&\nonumber\\
\int^\infty_0 dt \;
\frac{\sinh t}{\cosh^2\frac{t}{2}\cosh^2t}
\sin\frac{2t\theta}{\pi}&=&\frac{2\theta}{\cosh\theta}+\pi \sinh\theta\left(
\frac{1}{\cosh\theta +1}-\frac{1}{\cosh\theta}\right).
\end{eqnarray}
Combining, (\ref{Rbplusexpansion}) and (\ref{Zplusexpansion})
we deduce  expressions for the supersymmetric reflection factors
corresponding to the boundary conditions ${\cal BC}^+$
to order $\beta^2$:
\begin{eqnarray}\label{Kplusexpansion}
K^+_b\left( \theta \right) &\sim& \frac{i\sinh \theta+1}{i\sinh \theta-1}\,
M^+(\theta), \qquad K^+_f\left( \theta \right) \sim \frac{\cosh \theta}{i\sinh
\theta-1} \, M^+(\theta),\nonumber\\
&&\nonumber\\ M^+(\theta)&=&
 1-\frac{i\beta^2}{16\pi}\left( (2-\rho_0)\pi \sinh \theta \left(
\frac{1}{\cosh \theta+1}-\frac{1}{\cosh \theta} \right)
   +  \frac{2\rho_0\theta}{\cosh\theta}\right).
\end{eqnarray}
In a similar manner, the expansions of the reflection factors corresponding to
the boundary conditions ${\cal BC}^-$ are:
\begin{eqnarray}\label{Kminusexpansion}
K^-_b\left( \theta \right) &\sim& \frac{i\sinh \theta-1}{i\sinh \theta+1} \,
M^-(\theta), \qquad K^-_f\left( \theta \right) \sim \frac{i\sinh\theta
-1}{\cosh\theta} \, M^-(\theta), \nonumber\\ &&\nonumber\\ M^-(\theta)&=&
 1-\frac{i\beta^2}{16\pi}\left( (2-\rho_0)\pi \sinh \theta \left(
\frac{1}{\cosh \theta+1}-\frac{1}{\cosh \theta} \right)
   - \frac{2\rho_0\theta}{\cosh\theta}\right).
\end{eqnarray}

\section{Generating functional and two-point functions}
\setcounter{equation}{0}

Using a  path integral formalism and perturbation theory, one can obtain an
expression for the generating functional for the supersymmetric sinh-Gordon
model up to one-loop order. It is given by
\begin{eqnarray}
  Z =&& \bigg\{ 1 + \frac{i}{12}\beta^2 \int d^2x \left[ \theta(-x)m^2 \pm
      \frac{1}{4}\delta(x)m \right] \left[ 6 G^\pm(x,x) \left( \int d^2y
      \, G^\pm(x,y) J(y) \right)^2 \right. \nonumber\\
    &&\hspace*{7.1cm} - \left. \left( \int d^2y \, G^\pm(x,y)
J(y)\right)^4
      \right]\nonumber \\
    &&+\frac i4 m\beta^2 \int d^2x \theta(-x) \left[ -G^\pm(x,x) \int d^2y
      \, S^\pm_F(x,y) \eta(y)\int d^2z \ \bar \eta(z) \, S^\pm_F(z,x)
\phantom{\big(^2}
      \right. \nonumber\\
 && \hspace*{4cm} - S^\pm_F(x,x) \left( \int d^2y
   \ G^\pm(x,y) J(y) \right)^2 \nonumber \\
    &&+ \left. \left( \int d^2y \, G^\pm(x,y) J(y) \right)^2 \int d^2z \,
      S^\pm_F(x,z) \eta(z)\int d^2w \, \bar \eta(z) S^\pm_F(w,x) \right]
   \bigg\} Z_0. \label{3.36}
\end{eqnarray}

Using this we can proceed to evaluate up to the same order the boson and
fermion two-point functions which are defined by:
\begin{equation}
{\cal G}(x_1, x_2) = \left. \frac{1}{i^2} \frac{\delta^2Z}
{\delta J(x_1) \delta
    J(x_2)}\right|_{J = \eta = \bar{\eta} = 0},
\end{equation}
\begin{equation}
{\cal S}(x_1, x_2) = \left.\frac{1}{i^2} \frac{\delta^2Z}{\delta \eta(x_1)
\delta \bar \eta(x_2)}\right|_{J = \eta = \bar{\eta} = 0}.
\end{equation}
Using~(\ref{3.36}), we deduce an expression for the boson two-point function
in the form\footnote{\noindent For convenience we will denote  the propagators
$G^\pm(x_1,t_1;x_2,t_2)$ and $S^\pm_F(x_1,t_1;x_2,t_2)$ by $G^\pm(x_1,x_2)$
and $S^\pm_F(x_1,x_2)$, respectively.}
\begin{eqnarray}
{\cal G}(x_1, x_2) &=& G^\pm(x_1,x_2) - i\beta^2\int d^2x\,\left[
\theta(-x)m^2 \pm
      \frac{1}{4}\delta(x)m \right] G^\pm(x,x)
         \ G^\pm(x,x_1) \ G^\pm(x,x_2) \nonumber \\
   &&\hspace*{2cm}+\frac{im\beta^2}{2} \int d^2x \ \theta(-x) S^\pm(x,x)\
            G^\pm(x,x_1) \ G^\pm(x,x_2).
\end{eqnarray}
This are represented conveniently by means of the Feynman diagrams in Figure
1. However, it must be remembered that the diagrams are not the usual momentum
space diagrams. \vskip .5cm
\begin{figure}[h]
   \begin{center}
   {\epsfig{file=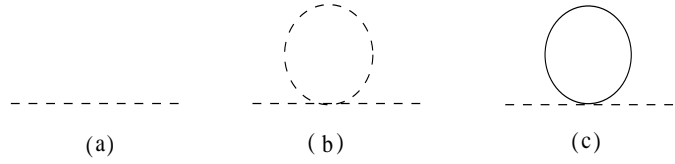,height=2cm, width=9cm, angle=0}}
  \end{center}
  \caption{Correction to the boson propagator.}
\end{figure}
\noindent Diagram (a) is the tree level boson propagator and the diagrams (b)
and (c) represent the boson and fermion one loop corrections to the boson two
point function respectively.

The fermion two-point function is given similarly by
\begin{equation}
{\cal S}(x_1, x_2) = S^\pm_F(x_1,x_2)- \frac{i}{4}m\beta^2 \int d^2x \
\theta(-x) G^\pm(x,x) \ S^\pm_F(x_1,x)  \ S^\pm_F(x,x_2).
\end{equation}
which may be  represented similarly by the Feynman diagrams of Figure 2.
\vskip .5cm
\begin{figure}[h]
    \begin{center}
       {\epsfig{file=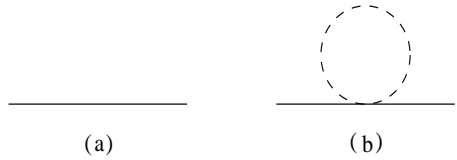,height=2cm, width=6cm, angle=0}}
    \end{center}
  \caption{Correction to the fermion propagator.}
\end{figure}

\section{The fermion reflection factors}
\setcounter{equation}{0}

In this section, we first calculate the fermion reflection factor
corresponding to the case ${\cal BC}^+$ when the boundary condition is
$\psi_1$ = $\psi_2$.

Since there is no additional boson-fermion coupling arising from the boundary
terms in (\ref{susyLag}), the correction to the fermion propagator will come
from the bulk coupling. This contribution corresponds to the diagram Fig~2(b)
and in detail it is:
\begin{equation}
  -\frac{i}{4}m\beta^2
    \int_{-\infty}^{+\infty} \hspace*{-4mm} dt''
    \int_{-\infty}^{0} \hspace*{-4mm} dx'' \;
 S_F^+\left( x, t;\ x'', t'' \right) \,
    G^+\left( x'', t'';\ x'', t'' \right) \,
    S_F^+ \left( x'', t''; \ x', t' \right), \label{expression}
\end{equation}
with the fermion propagator  given by~(\ref{fermion1}). The loop corresponds
to the integral
\begin{equation}
 \int \frac{d\omega''}{2\pi} \int\frac{dk''}{2\pi}\frac{i}
{\omega''^2 -k''^2 -m^2 +i\epsilon} \left[1+ K_b^+(k'')e^{-2ik''x''} \right].
\end{equation}
which needs a counter-term to remove the divergence. The energy integral is
finite  for the second term, and the finite part of the loop integral will be
\begin{equation}
\int \frac{dk''}{2\pi}\frac{1}{2\sqrt{k''^2+m^2}}
K_b^+(k'')e^{-2ik''x''}.
\end{equation}
Inserting this together with the fermion propagators in
expression~(\ref{expression}), we have
\begin{eqnarray}
  && -\frac{i}{8} m\beta^2  \int\frac{d\omega}{2\pi}\frac{dk}{2\pi}
\frac{dk'}{2\pi}\frac{dk''}{2\pi} \;
   \frac{ie^{-i\omega(t-t')}}{\omega^2-k^2-m^2+i\epsilon} \;
  \frac{ie^{-ikx-ik'x'}}{\omega^2-k'^2-m^2+i\epsilon}
  \frac{K_b^+(k'')}{\sqrt{k''^2+m^2}}\nonumber\\
  &&\nonumber\\
    &&\ \ \ \int_{-\infty}^0 dx''
 \left[ {\cal P}(\omega,k,k^\prime)\left(
  e^{ix''(k+k'-2k'')}+
    K_f^+(k)K_f^+(k')e^{ix''(-k-k'-2k'')}\right)\right. \nonumber \\
&&\nonumber\\
  &&\ \ \ \ \ \ \ \ \ \left. +{\cal Q}(\omega,k,k^\prime)
\left(K_f^+(k')e^{ix''(k-k'-2k'')}+
    K_f^+(k)e^{ix''(-k+k'-2k'')}\right)\right],
\end{eqnarray}
where
\begin{eqnarray}
&{\cal P}(\omega,k,k^\prime)=&\left(
    \begin{array}{cc} m^2+(\omega-k)(\omega-k') & -2im\omega+im(k-k')
    \\ 2im\omega+im(k-k') & m^2+(\omega+k)(\omega+k') \\
    \end{array} \right)\nonumber\\
    &&\nonumber\\
&{\cal Q}(\omega,k,k^\prime)=&\left( \begin{array}{cc} 2m\omega-m(k+k') &
-m^2i-i(\omega-k)(\omega+k')
    \\ m^2i+i(\omega+k)(\omega-k') & 2m\omega+m(k+k') \\
    \end{array} \right).
\end{eqnarray}
The next step is to perform the $x''$ integrals by using the following device
in which we set
\begin{equation}
  \int_{-\infty}^{0}dx'' e^{(ik+\sigma )x''} = \frac{-i}{k-i\sigma}, \label{vm}
\end{equation}
where $\sigma$ is a small positive constant, and take the limit
$\sigma\rightarrow 0$ at the end of the calculation. Using this we have
\begin{eqnarray}\label{kdoubleprime}
  && -\frac{i}{4} m\beta^2  \int\frac{d\omega}{2\pi}
    \int\frac{dk}{2\pi}\int\frac{dk'}{2\pi} \;
   \frac{ie^{-i\omega(t-t')}}{\omega^2-k^2-m^2+i\epsilon} \;
  \frac{ie^{-ikx-ik'x'}}{\omega^2-k'^2-m^2+i
    \epsilon}\; \int\frac{dk''}{2\pi}\frac{K_b^+(k'')}
     {2\sqrt{k''^2+m^2}}\nonumber \\
  && \left[ \left(
    \begin{array}{cc} m^2+(\omega-k)(\omega-k') & -2im\omega+im(k-k')
    \\ 2im\omega+im(k-k') & m^2+(\omega+k)(\omega+k') \\
    \end{array} \right)\frac{-i}{k+k'-2k''-i\sigma} \right.\nonumber  \\
  && +\left( \begin{array}{cc} 2m\omega-m(k+k') &
-m^2i-i(\omega-k)(\omega+k')
    \\ m^2i+i(\omega+k)(\omega-k') & 2m\omega+m(k+k') \\
    \end{array} \right)\frac{-iK_f^+(k')}{k-k'-2k''-i\sigma}\nonumber \\
  &&+ \left( \begin{array}{cc} 2m\omega-m(k+k') &
-im^2-i(\omega-k)(\omega+k')
    \\ im^2+i(\omega+k)(\omega-k') & 2m\omega+m(k+k') \\
    \end{array} \right)  \frac{K_f^+(k)}{k-k'+2k''+i\sigma}\nonumber \\
  && +\left. \left( \begin{array}{cc} m^2+(\omega-k)(\omega-k') &
    -2im\omega+im(k-k') \\ 2im\omega+im(k-k') & m^2+(\omega+k)(\omega+k')
\\
    \end{array} \right)\frac{iK_f^+(k)K_f^+(k')}{k+k'+2k''+i\sigma}
\right].
\end{eqnarray}
We can integrate out every  $k''$ integral by closing its contour in the upper
half plane and placing the branch cuts from $im$ to $i\infty$. This way, we
can avoid the pole contribution from $K^+_b(k'')$. The other  poles, involving
$\sigma$, are also avoided because $\sigma>0$.

Consider each term of  the $k''$ integral in turn starting with the first. It
can be decomposed into partial fractions and reexpressed as an integral over
the cut on the imaginary axis,
\begin{eqnarray}
 && \int\frac{dk''}{2\pi}\, \frac{K_b^+(k'')}{2\sqrt{k''^2+m^2}}
     \, \frac{-i}{k+k'-2k''-i\sigma}\nonumber \\
    &&\hspace{1.5cm} = \frac{1}{4\pi}\int_m^{\infty} dy \frac{1}{\sqrt{y^2-m^2}}
      \left[ \frac{1-K_b^+\left( \frac{k+k'}{2} \right)}{y+m} +
\frac{K_b^+
      \left(\frac{k+k'}{2} \right)}{y+i(k+k')/2} \right].
\end{eqnarray}
 Then, these  integrals can be evaluated
using the useful formula
\begin{equation}
   \int_{m}^{+\infty}dy \frac{1}{\sqrt{y^2 - m^2}}\frac{1}{y+2a}=
   \frac{2}{\sqrt{m^2 - 4a^2}}\left(
   \frac{\pi}{2}-\tan^{-1}\sqrt{\frac{m+2a}{m-2a}}\right).
\end{equation}
By changing the variable again via $y=m\cosh\theta$, we obtain the result for
the first $k''$-integral in (\ref{kdoubleprime}):
\begin{equation}
   \frac{1}{4\pi}\left[ \frac{1-K_b^+\left( \frac{k+k'}{2}\right)}{m} +
        \frac{2K_b^+\left( \frac{k+k'}{2}\right)}{\sqrt{m^2+(k+k')^2/4}}
       \left( \frac{\pi}{2} -
\tan^{-1}\sqrt{\frac{2m+i(k+k')}{2m-i(k+k')}}
     \right) \right].
\end{equation}
Having obtained the first $k''$ integral in~(\ref{kdoubleprime}), we now
perform the integrals over $k$ and $k'$. For these, the contours should be
closed in the upper half plane (because $x, x' < 0$) to yield
\begin{equation}
 \frac{1}{4\pi}e^{-i\hat k(x+x')} \frac{\omega}{2\hat k^2}
   \left( \begin{array}{cc} \omega-\hat k & -im
             \\ im & \omega+\hat k \\ \end{array} \right)
\left[\frac{1-K_b^+(\hat k)}{m} + \frac{K_b^+(\hat k)}{\cosh\theta}
\left(\frac{\pi}{4} -\frac{i\theta}{2} \right)\right],
\end{equation}
where $\hat k = \sqrt{\omega^2 - m^2}$.

The remaining three terms in~(\ref{kdoubleprime}) can be computed in a similar
fashion, except that $k+k'$ is replaced by one of  $k-k'$, $-k+k'$ and
$-k-k'$. Combining the results of all these calculations we have
\begin{eqnarray}
  -\frac{i}{16\pi}m\beta^2\int\frac{d\omega}{2\pi}e^{-i\omega(t-t')}
   e^{-ik(x-x')} \frac{1}{2\hat k}
   \left( \begin{array}{cc} \omega-\hat k & -2i
             \\ 2i & \omega+\hat k \\ \end{array} \right)
   \frac{K_f^+(\hat k)}{2m\sinh\theta} \nonumber\\
   \left[ 2\pi\sinh^2\theta \left(\frac{1}{\cosh\theta+1}-
 \frac{1}{\cosh\theta} \right)+\frac{4\theta\sinh\theta}{\cosh\theta}
\right].
\end{eqnarray}
\phantom{}From this we  extract the fermion reflection factor by selecting the
coefficient of the reflected term of the two-point function in the asymptotic
region far away from the boundary. Thus, in detail we find
\begin{equation}\label{Kplusdiagram}
 K_f^+(\hat k)=K_f^+(\hat k)_{class}\left[ 1-\frac{i\beta^2}{16\pi}\sinh\theta
  \left( \frac{1}{\cosh\theta+1}-\frac{1}{\cosh\theta} \right)
  -\frac{i\beta^2}{8\pi}\frac{\theta}{\cosh\theta} \right].
\end{equation}
This agrees precisely with~(\ref{Kplusexpansion}) provided we take
$\rho_0=1$ which is entirely consistent with $\rho(\beta)=B/4$.

The other fermion reflection factor, corresponding to
the boundary condition $\psi_1$ = -$\psi_2$, can be calculated to
the same order in a similar manner to obtain
\begin{equation}\label{Kminusdiagram}
 R_f^-(\hat k)= K_f^-(\hat k)_{class}\left[1-\frac{i\beta^2}{16\pi}\sinh\theta
   \left( \frac{1}{\cosh\theta+1}-\frac{1}{\cosh\theta} \right)
   +\frac{i \beta^2}{8\pi}\frac{\theta}{\cosh\theta} \right].
\end{equation}
The  calculations leading to (\ref{Kminusdiagram}) must be performed carefully
in view of the potential difficulties with the bound-state poles we alluded to
earlier. These problems may be circumvented by a judicious choice of contour,
picking up the poles related to $\sigma$ and carefully taking the limit
$\sigma\rightarrow 0$ at the end of the calculation. The expression
(\ref{Kminusdiagram}) also agrees with the expression for the fermionic
reflection factor corresponding to the boundary conditions ${\cal {BC}^-}$
which was quoted in (\ref{Kminusexpansion}).

\section{Conclusion}

In this paper, we have studied the boundary fermion  reflection factors for
the supersymmetric sinh-Gordon model perturbatively up to one loop. It is
gratifying that the results are in agreement with various conjectures obtained
on general grounds but disappointing that the calculations so far fail to
distinguish between the two favoured proposals given in (\ref{Rbplus}),
(\ref{Rbminus}), (\ref{MSRbplus}) and (\ref{MSRbminus}). Similar calculations
may be made to check the boson reflection factors but there is a stumbling
block which we have so far failed to overcome. In order to perform any
calculation of this kind we need to remove infinite parts of loop integrals
and this turns out to be straightforward for boson loops, as we have
described, but much less clear for the fermion loops (which contain more
divergences). In fact, at the moment it is not even clear to us that there is
a regularization scheme which will maintain the supersymmetry manifest in the
classical model. We shall return to this important question on another
occasion but the heart of the matter appears to be a need to make subtractions
which do not correspond to terms occurring in the original Lagrangian density,
especially at the boundary. If this is really the case, then the deductions
about the reflection factors made on general grounds using supersymmetry could
be suspect. Another approach, along the lines proposed in \cite{Corrigan2}, if
it can be developed supersymmetrically, should give information on the
boundary bound states which are present in the case ${\cal BC}^-$. In
particular, we are interested in knowing if these states have energies which
`float' with the coupling $\beta$. Basic questions concerning the model may
also be approached by alternative means, such as the Thermodynamic Bethe
Ansatz, which will require a knowledge of the reflection factors as an input
(see, for example a recent article by Ahn and Nepomechie \cite{Ahn}.)

\vspace{5mm}

\noindent {\large{\bf Acknowledgement}}\vspace{0.3cm}

\noindent One of us (MA) is grateful to the University of Durham for a Studentship.

\end{document}